\documentclass[aps,pra,twocolumn,superscriptaddress]{revtex4-2}

\usepackage{amssymb}
\usepackage{amsmath}
\usepackage{color}
\usepackage{graphicx}
\usepackage{hyperref}

\begin{document}

\title{Time reversal of complex evolution on a quantum computer}

\author{Mahdi Kourehpaz}
\affiliation{Mindbreeze GmbH, 4020 Linz, Austria}
\affiliation{Basic Research Community for Physics e.V., Germany}

\author{Dima L. Shepelyansky}
\email{dima@irsamc.ups-tlse.fr}
\affiliation{Univ Toulouse, CNRS, Laboratoire de Physique Th\'eorique, Toulouse, France}

\date{\today}

\begin{abstract}
The Boltzmann-Loschmidt dispute in 1876-1877 discussed the problem of time reversal of thermalization from time reversible classical equations of motion. Here, 150 years later, we highlight this problem in the frame of quantum computing. A quantum protocol is proposed that allows to perform a time reversal of complex evolution in the regime of many-body quantum chaos with many qubits. The system represents an evolution of qubits on a square lattice with inter-qubit next-nearest static couplings with a driven pulsed magnetic field. The system entropy grows rapidly to maximal values but returns to initial small values after time reversal. This time reversal is shown to be stable with respect to quantum gate imperfections. However, similar to the Lorenz butterfly effect, there is the butterfly effect of qubit when inversion of only one qubit breaks time reversibility of the whole system. It is argued that this protocol is accessible to nowadays quantum computers and annealers with hundreds of qubits.
\end{abstract}

\maketitle

\section{Introduction}
\label{sec1}

The statistical theory of entropy growth and thermalization 
had been formulated by Boltzmann in 1872 on the basis of dynamical equations of motion of
atomic gas \cite{boltzmann1}. This theory was objected to by Loschmidt in 1876 \cite{loschmidt}
who pointed out that the reversible dynamical equations
allow to invert velocities of atoms returning them back to the initial state.
The reply of Boltzmann was published in 1877 \cite{boltzmann2}
and the legend tells that on the direct question on
what happens with entropy and thermalization if all atom velocities
are inverted he replied {\it then go and invert them} \cite{mayer}.

In the frame of classical mechanics the resolution
of Boltzmann-Loschmidt dispute is related to emergence of dynamical chaos
with its exponential instability of nonlinear motion that
leads to an exponential growth of errors thus breaking
time reversal on logarithmically short times
\cite{arnold,sinai,chirikov1979,lichtenberg}. An example
of such time reversal breaking
is given in \cite{dls1983}.

To illustrate such an exponential instability of chaotic motion
E. N. Lorenz in 1972 illustrated this by a flap of a butterfly's wings in Brazil
setting off a tornado in Texas \cite{lorenz1972}.
This became known as the butterfly effect \cite{butterflywiki}.

In the frame of quantum mechanics the time evolution is described by the linear 
Schr\"odinger equation and the Lyapunov exponential instability of motion (with exponent $\Lambda$)
exists only on a logarithmically short Ehrenfest time scale $t_E \sim |\ln \hbar |/\Lambda $ 
after which mixing stops at a small phase space scale determined by the Planck constant $\hbar$
(taken here in dimensionless units)
\cite{chirikov1981,dls1981,chirikov1986,haake,dlsehrenf}.
Hence for a quantum chaos evolution the time reversal is preserved at
a moderate strength of computer errors
while the corresponding classical evolution is
nonreversible on the same computer
even at tiny round-off computer errors
\cite{dls1983}.
The effects of errors and perturbations on time reversal of evolution
in the regime of quantum chaos, known as the Loschmidt echo,
have been studied by different groups
with detailed description given in
\cite{peres,jalabert1,frahm,prosen,jacquod,jalabert2}.

The experimental realization of time reversal was first
performed in spin systems as spin echo \cite{hahn1,pastaw,hahn2}.
For acoustic and electromagnetic waves time reversal was realized in
\cite{fink1,fink2} leading to important useful applications.
Time reversal technique is also broadly used for
a seismic analysis in geophysics \cite{geopht}.
For atomic Bose-Einstein condensate a time reversal of a part of wave
packet was achieved in experiments with 
atoms in kicked optical lattice \cite{bec1,bec2}
following the theoretical proposal \cite{qbl1}.
An experiment with time reversal of almost all wave packet of cold atoms, or ions,
in a harmonic trap was recently proposed in \cite{bl150}.

In recent decades, there has been a significant progress in the development
of quantum computation and information (see e.g. \cite{steane,chuang,deutsch}). Quantum computing now is done with several tens
of qubits and even with more than hundreds of qubits
(see e.g. \cite{smela,smela2,91qu,144qu,74qu}).
Even though a significant number of qubits has been reached now in
quantum computing, still the realization of several thousands of
two-qubit gates remains a challenge 
due to decoherence effects. For certain systems
it is easier to realize quantum simulators of specific 
physical effects like e.g the time evolution
of Ising Hamiltonians or kicked Ising chains (see e.g. \cite{91qu,144qu,74qu}).

It is important to note that many interesting
mathematical and numerical results
for kicked Ising chains
have been obtained in the regime of quantum chaos
(see \cite{prosen1,prosen2,prosen3,prosen4} and Refs. therein).
The quantum simulations of such kicked Ising chains
in certain respects are more accessible
for nowadays quantum computers or simulators (see e.g. \cite{91qu,144qu}).

While in \cite{prosen1,prosen2,prosen3,prosen4} the
studies are mainly done for kicked Ising chains
here we consider a quantum computer
with qubits forming a two-dimensional (2D) square lattice.
Such quantum computer hardware models
with static imperfections in 2D
have been studied e.g. in \cite{qch1,qch2,nobel}.
Here we show that a specific case of kicked Ising 2D model
allows to realize efficiently a complex evolution
in a regime of quantum chaos when a rapid entropy growth
is inverted in time with full recovering of the initial quantum register state.
We argue that such type of quantum computation
on a quantum computer allows to highlight the Boltzmann-Loschmidt dispute
in a new frame of quantum reality.

We also point out that the proposed TIme Reversal Kicked Ising (TIRKI) model
can be also realized on quantum annealers
that can operate with thousands of qubits
solving difficult physical problems (see e.g. \cite{qa1,qa2,qa3} and Refs. therein).
The time reversal in a system with several hundreds of qubits
would allow to realize the inversion of time arrow
for a macroscopically huge number of Hilbert states.

The paper is organized as follows:
the TIRKI model is presented in Section 2,
the properties of many-body quantum chaos in this model
are described in Section 3,
the results for time reversal in the model
are described in Section 4,
the effects of static imperfections on
time reversal are considered in Section 5,
the butterfly effect of qubit
for time reversal is analyzed in Section 6,
and discussion of the results is given in Section 7.

\section{Description of TIRKI model}
\label{sec2}

The TIRKI model of a driven quantum computer is described by the Hamiltonian:
\begin{equation}
  \label{hamil}
  H =  H_0 + V\delta(t-T) = J\sum_{i,j}  \sigma_{i}^z \sigma_{j}^z
  +  h \sum_{i}\sigma_{i}^x \times \sum_m \delta(t-mT) \; ,
\end{equation}
where $H_0$ is unperturbed Hamiltonian and $V$ describes a time-dependent kick
perturbation with a periodic $\delta$-function of period $T$;
$\sigma^x$ and $\sigma^z$ are Pauli matrices of qubits,
sum over $i,j$ runs over next-nearest neighbor qubits located on
a square lattice with hard boundary conditions; $J$ and $h$ describe the
qubit coupling strength and the kick amplitude. The number of qubits is $n_q=L_x \times L_y$
with lattice sizes $L_x,L_y$ in $x,y$-directions. The size of the Hilbert space
is $N = 2^{n_q}$ and the quantum register states are $|b_i\rangle =|10\cdots 1\rangle$
with $0$ or $1$ for spin down or up.

The time evolution propagator of wavefunction $\psi$ of (\ref{hamil})
on one period $T$ has the form:
\begin{equation}
  \label{evol}
  \bar{\psi} = U\psi = U_H(\varepsilon) U_K(h) \psi = \exp(-i \varepsilon \sum_{i,j}
  \sigma_{i}^z \sigma_{j}^z) \exp(- i h\sum_{i}\sigma_{i}^x ) \psi ,
\end{equation}
where $\bar{\psi}$ is the wavefunction after one perturbation period and $\varepsilon = JT$.
The unitary propagator $U_K$ can be considered as a pulsed magnetic field acting 
simultaneously on all qubits
while the operator $U_H$ simply describes a free evolution of coupled qubits.
The quasienergy eigenfunctions $\psi_\lambda$ of the evolution operator $U$ are
determined by the equation $U\psi_\lambda = \lambda \psi_\lambda$
with $|\lambda|=1$.

The time reversal protocol
is the following: the evolution is done during
$t_r$ periods (kicks) with the propagator (\ref{evol});
after that the time reversal evolution
operator takes the form
\begin{equation}
  \label{reversal}
  U_r = U_K(-h) U_H(T') = {U_K(h)}^* {U_H(\varepsilon)}^* = U^\dag \; ,
\end{equation}
where $U_K(-h)$ is the kick operator with $h$ replaced by $-h$
(magnetic field changes sign); and the period between kicks
being $T' = (2\pi - \varepsilon)/J$ so that we have
$\varepsilon \rightarrow 2\pi - \varepsilon$.
However, at $\varepsilon =2\pi$ the operator $U_K(2\pi) = I$
where $I$ is unit operator. As a result
we obtain complex conjugation operator
and after a back preparation of $t_r$
periods of $U_r=U^\dag$ the whole system
returns to the initial state
at the time moment $t=2t_r$.
The same time reversal protocol
works also for any
unperturbed Hamiltonian $H_0$
being an integer function of ${\sigma_i}^z, {\sigma_j}^z$
multiplied by $J$
(so that eigenvalues of $H$ are integers multiplied by $J$).
For example this can be the unperturbed Hamiltonian
$H_0 = J (\sum_{i,j}  \sigma_{i}^z \sigma_{j}^z + A \sum_i \sigma_{i}^z)$
where $A$ is any integer.

Thus the time reversal of quantum evolution
(\ref{evol}) is rather simple for realization:
one simply needs to replace $\varepsilon \rightarrow 2\pi - \varepsilon$
by increasing the time interval between kicks
and by changing sign of pulsed magnetic field $h \rightarrow -h$.

To study how robust is the phenomenon of time reversal
we consider how it is influenced by
static imperfections following the approach used in \cite{frahm}.
With imperfections the Hamiltonian takes the form:

\begin{equation}
  \label{hamil2}
  H =  H_0 + V\delta(t-T) = \sum_{i,j} J_{ij} \sigma_{i}^z \sigma_{j}^z
  +  \sum_{i} h_i\sigma_{i}^x \times \sum_m \delta(t-mT) \; ,
\end{equation}
with $J_{ij} =J +\eta_{ij}$ and $h_i= h +\mu_i$
where random static perturbations are distributed in narrow intervals
$-\eta/2 \leq \eta_{ij} \leq \eta/2$
(they act only for nearest-neighbor qubits as in (\ref{hamil}))
and $-\delta_x/2 \leq \mu_i \leq \delta_x/2$. 
Thus with imperfections we have the evolution operator for $t \leq t_r$:
\begin{equation}
  \label{evolim}
   U = U_H(\varepsilon) U_K(h) \psi = \exp(-i  \sum_{i,j}
   (\varepsilon +\varepsilon_{ij})\sigma_{i}^z \sigma_{j}^z)
   \exp(- i \sum_{i}(h+\mu_i) \sigma_{i}^x ) \psi ,
\end{equation}
where amplitudes of weak random imperfections are
distributed in narrow intervals
$-\delta_z/2 \leq \varepsilon_{ij} =T \eta_{ij} \leq \delta_z/2$ and
$-\delta_x/2 \leq \mu_i \leq \delta_x/2$.

\begin{figure}
\begin{center}
\includegraphics[width=0.85\columnwidth]{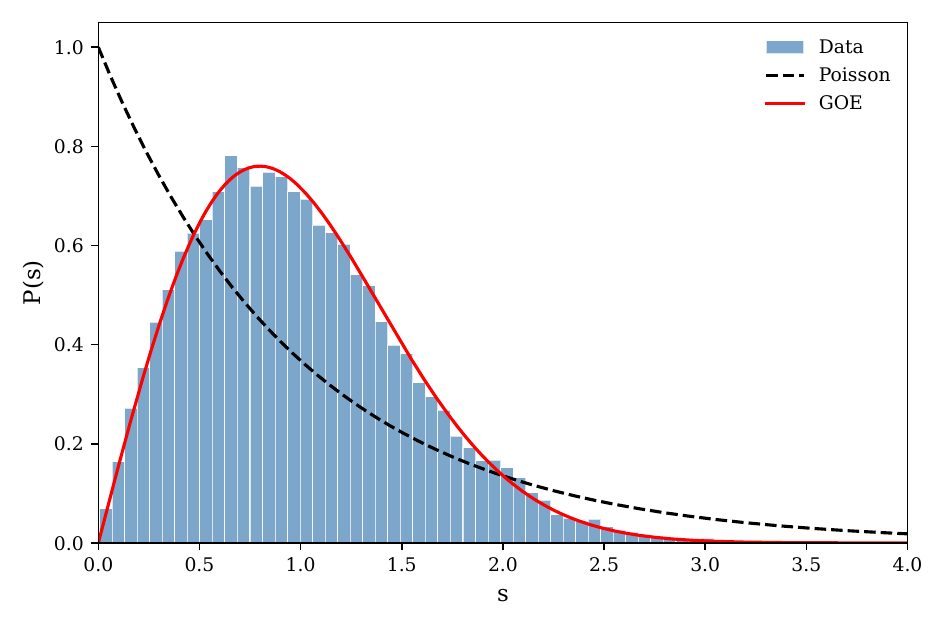}
\end{center}
\caption{\label{fig1}
  Level spacing statistics of quasienergies $\lambda$ of unitary operator $U$ at
  parameters $\varepsilon =2$, $h=1$, $n_q=4\times 4 =16$; it is 
  obtained by diagonalizing all ten symmetry sectors separately,
  there are approximately $4.9\times 10^4$ spacings. The Poisson (dashed curve) and
  Wigner surmise of RMT (red solid curve) distributions are also shown.
}
\end{figure}

For the time reversal part $t_ r < t \leq 2t_r$
we have $T \rightarrow (2\pi -\varepsilon)/J$ and
$h \rightarrow -h$ so that the backward evolution propagator
with imperfections is:
\begin{equation}
  \label{revolim}
  U_r =  \exp(- i \sum_{i}(-h+\mu_i) \sigma_{i}^x )
  \exp(-i \sum_{i,j} (-\varepsilon + (2\pi-\varepsilon) \varepsilon_{ij})\sigma_{i}^z \sigma_{j}^z) \psi .
\end{equation}
Thus for $t_r < t \leq 2t_r$ the strength of imperfections is increased
for $\varepsilon < \pi$ with $\delta_z \rightarrow (2\pi -\varepsilon)\delta_z$.
We assume that $\mu_i$ do not change sign.

We characterize the precision of time reversal by
fidelity defined in a usual way as
$F(t) = |\langle\psi(t=0)|\psi(t)\rangle|^2$.
In absence of imperfections we have
exact time reversal and $F(t=2t_r) = 1$.
The fidelity $F(2t_r)$ decreases with increase of imperfections.

Also we study the level spacing statistics
of quasienergies $\lambda$. Following \cite{qch1,qch2,nobel} we
also characterize the eigenstates $\psi_\lambda$
by their von Neumann entropy defined as
$S_q = - \sum_i W_i \log_2 W_i$ where $W_i$ is the quantum probability
to find the noninteracting multiqubit
quantum register state $|\varphi_i\rangle$ of Hamiltonian $H_0$ in the eigenstate
$|\psi_\lambda \rangle$ of the evolution operator $U$ at $h >0$.
The same definition of $S_q$ can be used for a
evolution of wavefunction $\phi(t)$ at some time moment $t$
(instead of $|\psi_\lambda \rangle$).

We also note that without imperfections
the kick operator at $h=\pi/2$ is just a global spin-flip operator
$U_K(\pi/2) = \exp(-i \pi/2 \sum_i\sigma_i^x)=-i\Pi_i\sigma_i^x$ and
since the term $\sigma_i^z \sigma_j^z$ in the $U_H$ is invariant under a global spin flip,
$U_K(\pi/2)$ commutes with $H$.
As a result the characteristics of time evolution
have periodic dependence on $h$ with $\pi/2$ period (see below).
For a square lattice and without imperfections, the lattice has the symmetry
of the dihedral group $D_4$ with five irreducible symmetry sectors 
$A_1$, $A_2$, $B_1$, $B_2$, and $E$. In addition, the time evolution
operator $U$ is invariant under the global spin flip operator $[U, \Pi_i \sigma_i^x]=0$.
So in total, we have ten symmetry sectors $D_4\times\mathbb{Z}_2$. Finally,
$U$ is diagonalized exactly in each sector separately to calculate
the quasienergy spacings in Figure \ref{fig1}.

We note that a quantum time reversal protocol for
quantum computing of the classical chaotic Arnold cat map
had been discussed in \cite{arnold1,arnold2}.
However, this protocol
requires a significant number of two-qubit and one-qubit gates
and is still not accessible to nowadays quantum computers
(about $10^3$ gates for one map iteration at 27 qubits).
In contrast, the protocol proposed here
is easy for implementation both with quantum computers and quantum annealers:
for one evolution period one keeps fixed static inter-qubit couplings and
makes a short pulse of global magnetic field
performing fixed phase rotation
for all qubits.

\section{Many-body quantum chaos in TIRKI model}
\label{sec3}

Here we present results showing that in the absence of imperfections at
typical parameter values $\varepsilon \sim 1, h \sim 1$
the quantum register states of Hamiltonian $H_0$ (\ref{hamil})
are mixed by the kick potential $V$ and the system
is the regime of quantum chaos (see e.g. \cite{haake,qch1,qch2,nobel}).
Indeed, for the typical parameters $\varepsilon, h$ we show that the
level spacing statistics $P(s)$ of nearest quasienergies $\lambda$
is described by the Random Matrix Theory (RMT) \cite{haake} (see Fig.~\ref{fig1},
here $s$ is rescaled level spacing).
Indeed, the data shows that $P(s)$ nicely follows the Wigner surmise
$P_W(s) = (\pi s/2)\exp(-\pi s^2/4)$.
This justifies that the TIRKI model at such parameters is in the regime of
developed quantum chaos.

\begin{figure}
\begin{center}
  \includegraphics[width=0.45\columnwidth]{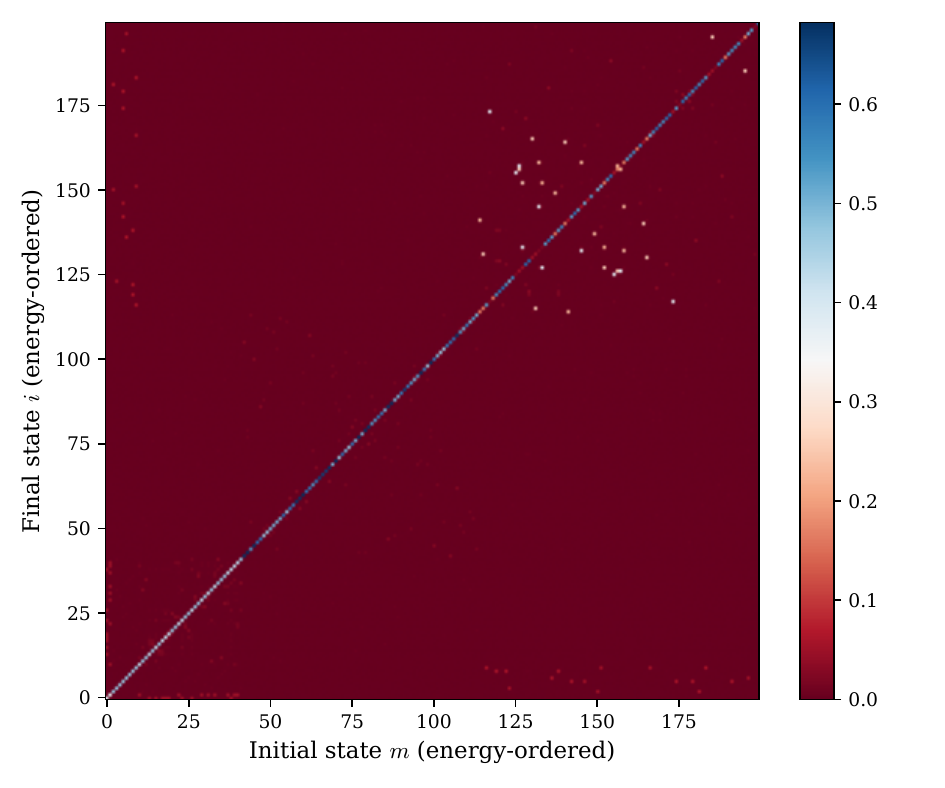}
  \includegraphics[width=0.45\columnwidth]{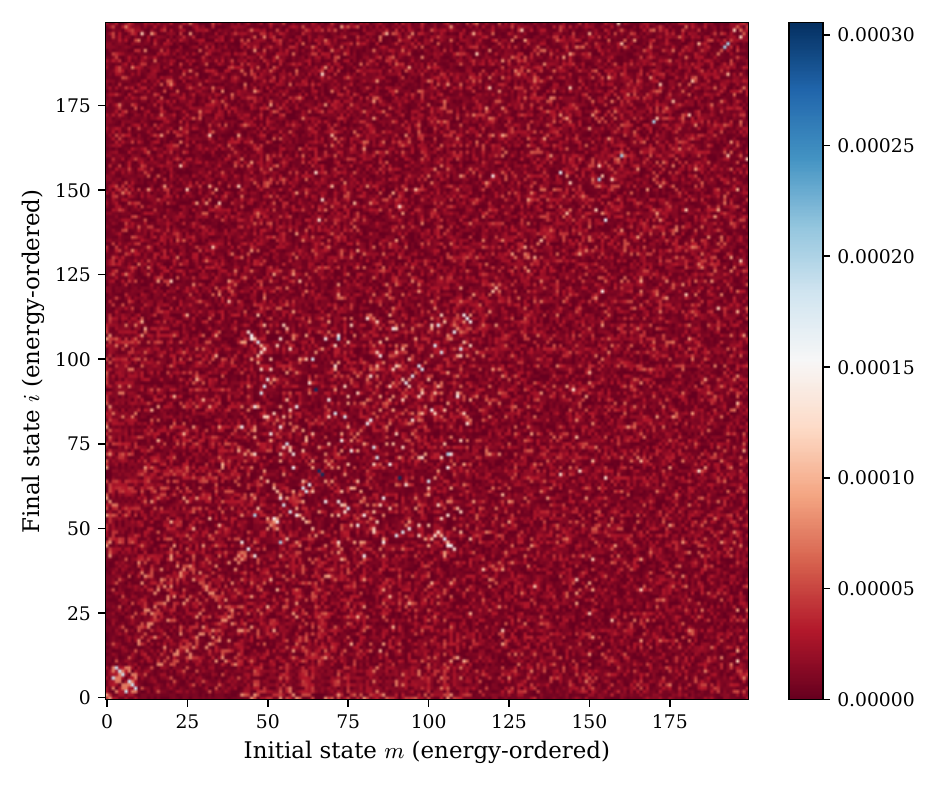}
\end{center}
\caption{\label{fig2}
  Matrix of excited state $W_{im}$ after $t_r=30$ perturbation periods
   at $n_q=4 \times 4$, $\varepsilon =2$, and 
   $h=0.1$ (left); $h=1$ (right);
   only 200 initial register states
   with minimal energy $E_0=H_0$ are shown;
   $m/i$ is index of initial/excited register state. 
}
\end{figure}

In Fig.~\ref{fig2} we show that after a time evolution during $t=30$
the initial register states ($200$ of them at lowest energy $E_0=H_0$ are shown)
the kick perturbation $V$ does not excite significant number of other register states at $h=0.1$
while for $h=1$ many of register states are excited. Thus at $h=0.1$ the system
(\ref{hamil}) is close to an integrable case while
at $h=1$ the system is in the regime of developed quantum
chaos in agreement with the RMT level spacing statistics shown in Fig.~\ref{fig1}.

\begin{figure}
\begin{center}
\includegraphics[width=0.85\columnwidth]{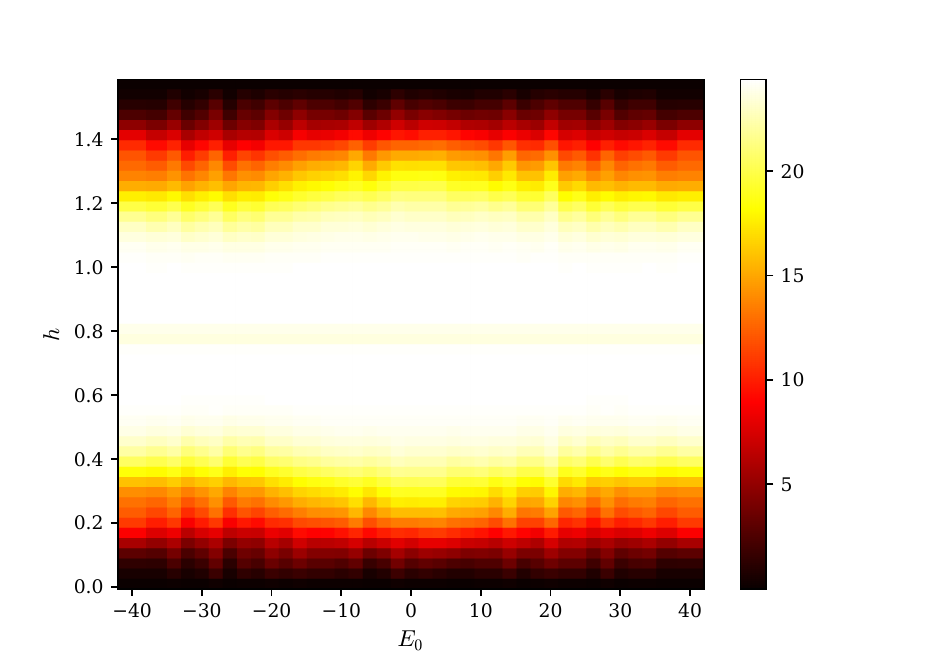}
\end{center}
\caption{\label{fig3} Quantum register melting: color bar shows entropy values $S_q$ after $t_r=30$
  evolution periods in dependence on $h$ and initial energy $E_0=H_0$
  for $n_q = 5 \times 5 = 25$ qubits, $\varepsilon=2$, $t_r=30$,
  $|\psi_0\rangle=|\uparrow \uparrow \uparrow \dots \uparrow \rangle$.
  Melting sets in above a threshold in $h$ and 
  is strongest in the high-density central band 
  $(E_0 \approx 0)$, invading the low-density band edges later; 
  the horizontal dark stripes are the fidelity register preserving areas.
}
\end{figure}

Another confirmation
of the emergence of quantum chaos is given in Fig.~\ref{fig3}
where we show the dependence of entropy $S_q$ after
$t_r=30$ kicks as a function of kick strength $0 < h < \pi/2$ and
initial register energy $E_0=H_0$ for $\varepsilon =2$ and $n_q = 5 \times 5 =25$.
Approximately, for weak values of $h < 0.1$ the entropy $S_q$ remains relatively
small $S_q < 2$ while for $h>0.5$ it reaches its maximal value $S_q \approx 25$.
Thus the system is in the regime of strong chaos for $h > 0.5$.
Also due to periodicity of kick
strength with $h=\pi/2$ the values of $S_q$ are periodic and symmetric in respect to
$h \rightarrow \pi/2 - h$.

The results of Figs.~\ref{fig1},~\ref{fig2},~\ref{fig3} show
that at typical parameter values $\varepsilon \sim h \sim 1$
the system (\ref{hamil}) is in the regime of developed quantum many-body chaos
characterized by complex evolution in time.

\section{Time reversal for TIRKI model}
\label{sec4}

Here we discuss the properties of time reversal
protocol described in Section~\ref{sec2}
(mainly in absence of imperfections).
The effects of imperfections are discussed in Section~\ref{sec5}.

\begin{figure}
\begin{center}
\includegraphics[width=0.85\columnwidth]{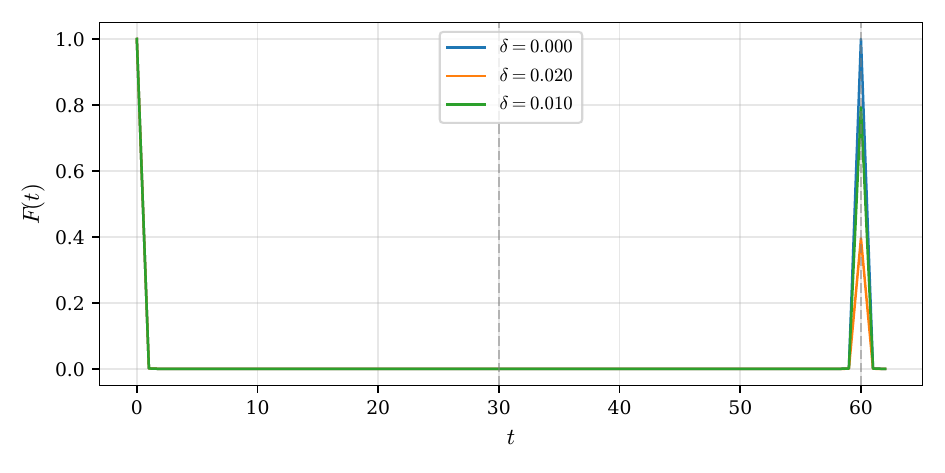}
\end{center}
\caption{\label{fig4}
  Fidelity $F(t)$ dependence on time $t$
  for model parameters $\varepsilon=2$, $h=1$, $n_q=5\times 5$,
  $t_r=30$ (vertical dashed line), initial state is
  $|\psi_0\rangle=|\uparrow \uparrow \uparrow \dots \uparrow \rangle$;
  amplitude of static imperfections is $\delta_z=\delta_x=\delta$
  with $\delta =0.02$ (orange curve); $0.01$ (green curve); $0$ (blue curve, no imperfections).
}
\end{figure}

In Fig.~\ref{fig4} we show the time dependence of fidelity $F(t)$ on time
 with the time reversal done at time moment $t=t_r=30$
at system parameters $\varepsilon=2$, $h=1$ and $n_q=5\times 5 =25$ qubits.
Already after one period the fidelity drops to a very small value
but at the return moment $t=2t_r=60$ it is perfectly recovered at $F(2t_r)=1$.
The effect of static imperfections is shown by the dashed curves at
$\delta_z=\delta_x= \delta$. These imperfections reduce the fidelity revival
at $t=2t_r$ but still the fidelity remains relatively high at $\delta=0.01$
(imperfections effects are discussed in detail in Section~\ref{sec5}). 

\begin{figure}
\begin{center}
\includegraphics[width=0.85\columnwidth]{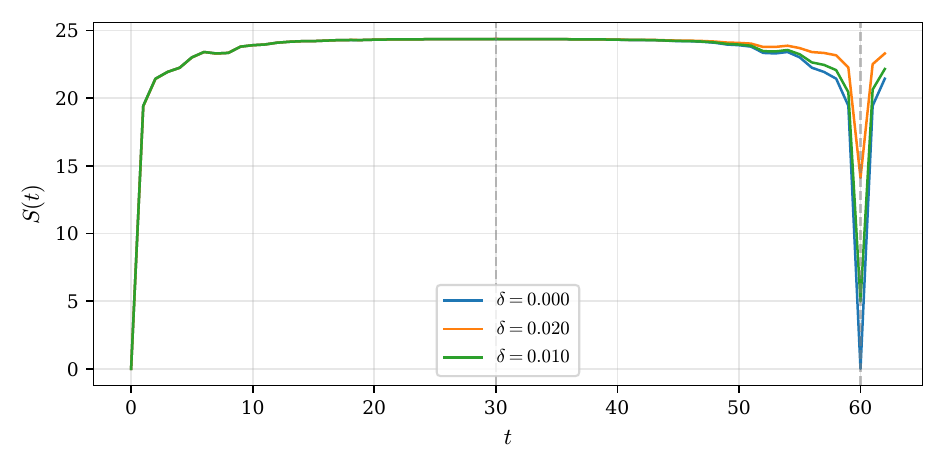}
\end{center}
\caption{\label{fig5}
  Entropy $S_q$ dependence on time $t$
  for the same system parameters as in Fig.~\ref{fig4}.
}
\end{figure}

The time reversal effect for entropy $S_q(t)$ in shown in Fig.~\ref{fig5}
for the same system parameters. After first 5 periods of time
the entropy grows from its initial value $S_q=0$ almost up to its maximal value $S_q =25$
remaining close to this value on longer times.
At the return time $t=2t_r=60$ the
entropy returns to its initial value $S_q=0$.
The backward return of entropy $S_q(2t_r)$ is only weakly
affected by static imperfections (see 
more detail in Section~\ref{sec5}).

\begin{figure}
\begin{center}
\includegraphics[width=0.85\columnwidth]{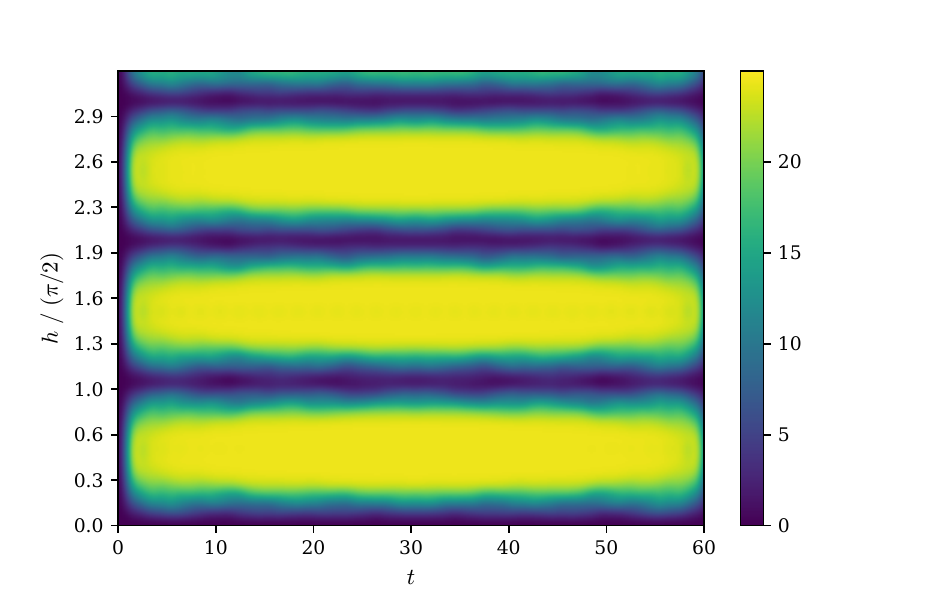}
\end{center}
\caption{\label{fig6}
  Entropy $S_q$ dependence on $h$ and time $t$
  shown by color (color bar with $S_q$ values); here $\varepsilon =2$,
  $n_q=5 \times 5 =25$,
  $|\psi_0\rangle=|\uparrow \uparrow \uparrow \dots \uparrow \rangle$,
  $t_r=30$ and no imperfections.
The kick strength expressed as $h/(\pi/2)$ (vertical). Bright (yellow) bands
mark fast saturation to a large entropy; the dark horizontal
stripes at $h/(\pi/2) = 0, 1, 2, 3$, are the register-preserving resonances
where the kick is trivial up to a phase.
}
\end{figure}

The dependence of $S_q(t)$ on time and parameter $h$ is shown in Fig.~\ref{fig6} at $\varepsilon=2$
for the time interval $0 \leq t \leq 2t_r =60$.
The results clearly show the time reversal of entropy $S_q(t)$.
As in Fig.~\ref{fig3} the entropy remains small $S_q < 2$ for $h < 0.1$
being in a quasi-integrable phase while
the phase of developed quantum chaos appears at $h > 0.1$
(as discussed above there is periodicity and symmetry for $h$ variation).

\section{Effects of imperfections for time reversal}
\label{sec5}

\begin{figure}
\begin{center}
\includegraphics[width=0.85\columnwidth]{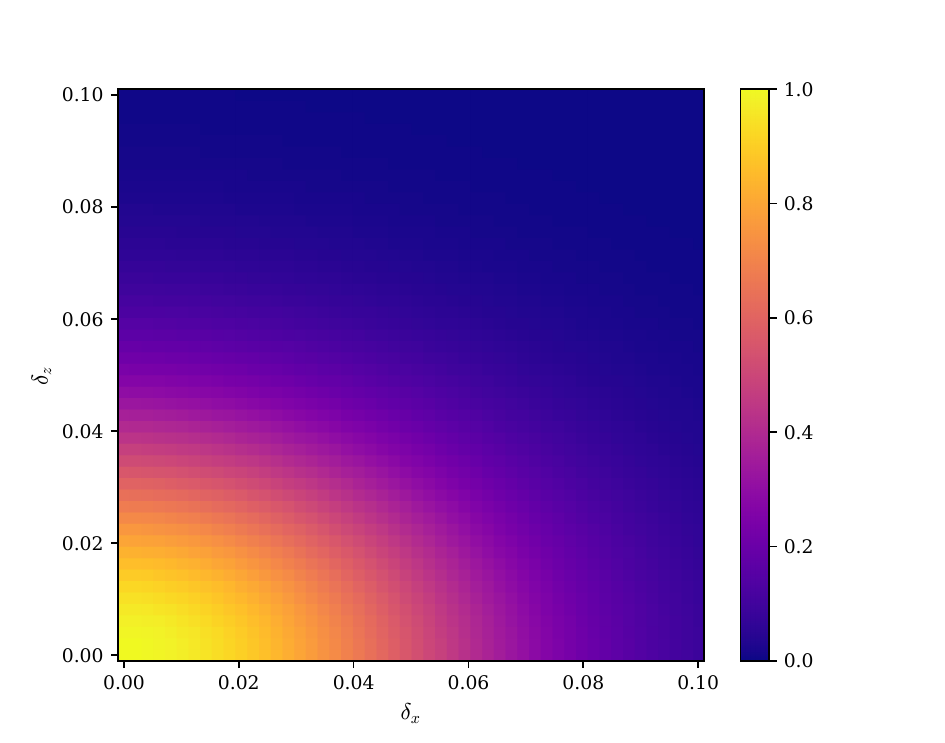}
\end{center}
\caption{\label{fig7}
  Fidelity $F(2t_r)$ shown by color bar vs. imperfection amplitudes $\delta_x$, $\delta_z$
  at $n_q=4\times 4 =16$, $\varepsilon =2$, $h=1$, $t_r=30$,
  initial state is $|\psi_0\rangle=|\uparrow \uparrow \uparrow \dots \uparrow \rangle$;
  $\delta_x$ is the spread of the kick angle per site, and
  $\delta_z$ is the spread of the Ising coupling per bond,
  data is for a single fixed realization.
  The reversal is exact only at $\delta_x=\delta_z=0$.
}
\end{figure}

The effects of static imperfections for fidelity $F(2t_r)$
are shown in Fig.~\ref{fig7} with its dependence on
imperfection amplitudes $\delta_x, \delta_z$.
The results show that the fidelity of time reversal
remains robust against
these imperfections.

\begin{figure}
\begin{center}
\includegraphics[width=0.85\columnwidth]{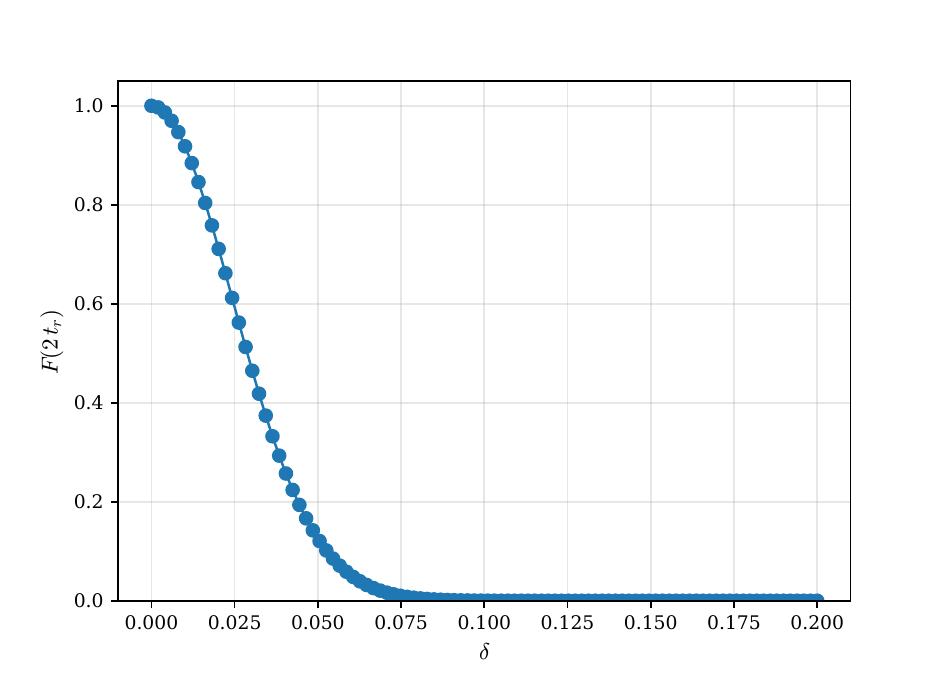}
\end{center}
\caption{\label{fig8}
  Fidelity $F(2t_r)$ vs $\delta_z=\delta_x=\delta$; here
  $\varepsilon=2$, $h=1$, $n_q=16$, $t_r=30$,
  $|\psi_0\rangle=|\uparrow \uparrow \uparrow \dots \uparrow \rangle$.
  This figure is a cut through the diagonal of Fig.~\ref{fig7}. 
}
\end{figure}

The dependence of fidelity at the return time $F(2t_r)$
on the amplitude of imperfections $\delta_z=\delta_x=\delta$
is shown in Fig.~\ref{fig8}. It is approximately described by the expression
$F(2t_r) \sim \exp(-G \delta^2 t_r)$ where $G$ depends on system
parameters $h, \varepsilon, n_q$ being in agreement with the general
properties of Loschmidt echo decay. The Loschmidt echo decay
is studied and described in detail in \cite{jalabert1,frahm,prosen,jacquod,jalabert2,qch1}
and due to that we do not discuss this behavior in detail here.

\begin{figure}
\begin{center}
\includegraphics[width=0.85\columnwidth]{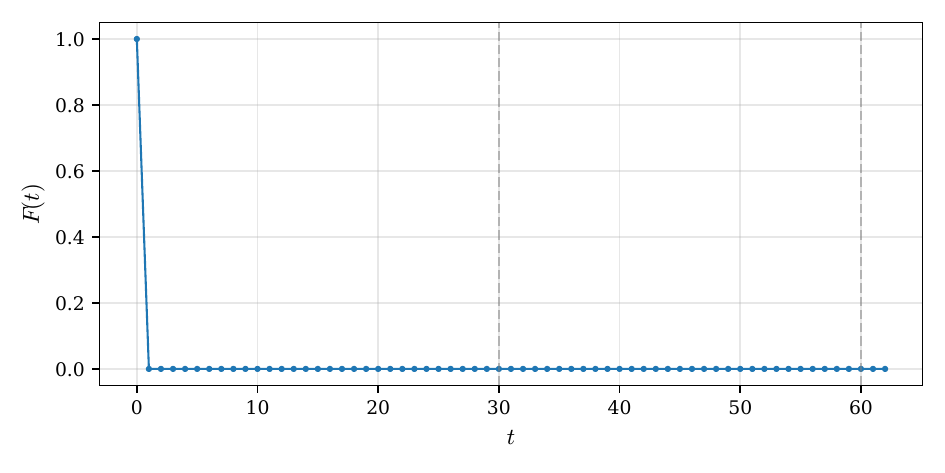}
\end{center}
\caption{\label{fig9}
  Fidelity $F(t)$ dependence on time $t$ for the butterfly effect of qubit: one central qubit is inverted
  at $t_r=30$. Here $n_q=5 \times 5 =25$, $\varepsilon=2$, $h=1$,
  $|\psi_0\rangle=|\uparrow \uparrow \uparrow \dots \uparrow \rangle$ and there are no imperfections.
}
\end{figure}

\begin{figure}
\begin{center}
\includegraphics[width=0.85\columnwidth]{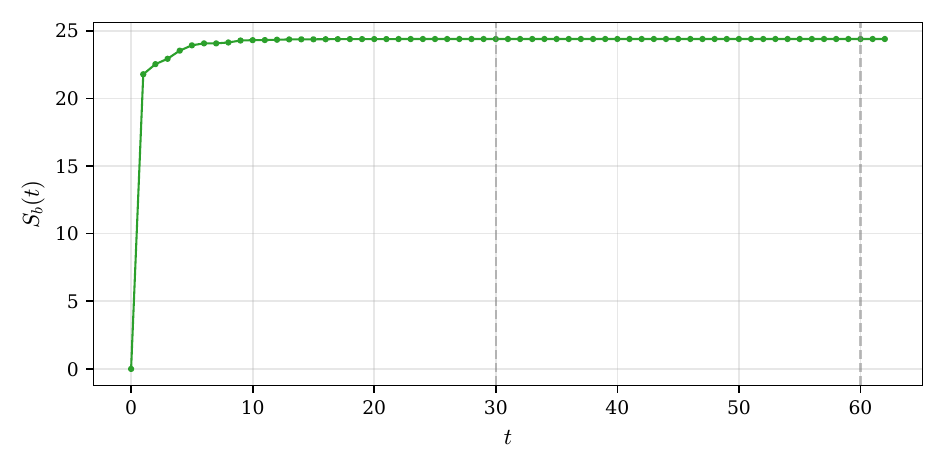}
\end{center}
\caption{\label{fig10}
  Same as in Fig.~\ref{fig9} but for system entropy $S_q(t)$ vs $t$.
}
\end{figure}

\section{Butterfly effect of qubit}
\label{sec6}

The above results show that quantum computing of complex evolution (\ref{evol})
remains stable with respect to quantum gate imperfections. This stability is also
preserved in respect to noise phase gate errors as it is shown e.g. in \cite{frahm}. 
However, it is possible to ask if the time reversal operation would remain
stable if at a time reversal moment $t_r$ there is inversion
of only one qubit (apart of this the time reversal protocol remains
unchanged). This time reversal operation can be viewed as the butterfly effect of qubit.
The butterfly effect in the classical world with classical equations of motion and
chaotic dynamics was discussed by E.N.Lorenz \cite{lorenz1972}
(see also the story written by R.Bradbury \cite{bradbury}).

\begin{figure}
\begin{center}
\includegraphics[width=0.85\columnwidth]{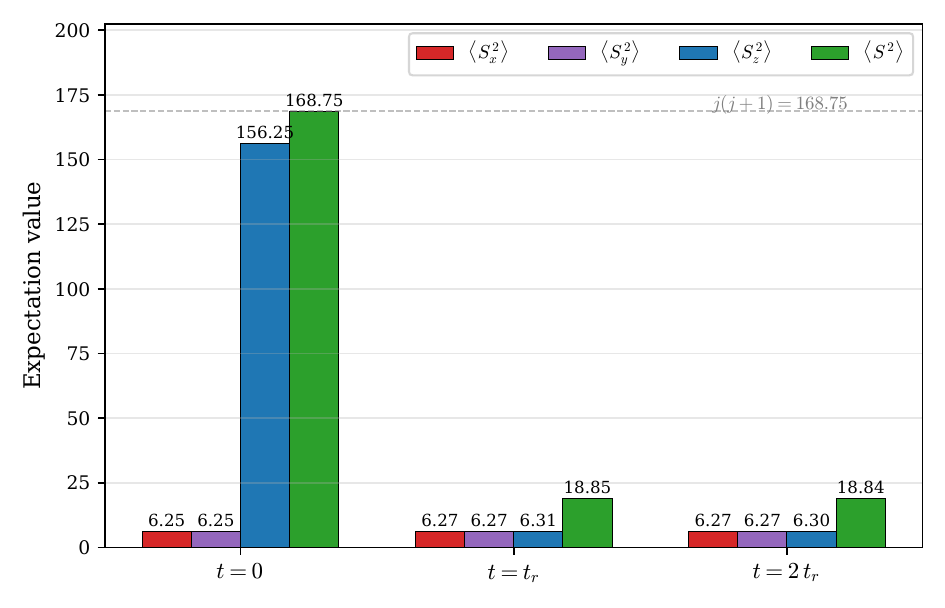}
\end{center}
\caption{\label{fig11}
  Spin square $S^2(t)$ of the whole system shown at time moments $t=0; t_r=30; 2t_r=60$
  for the butterfly effect of qubit at the same system parameters as in Fig.~\ref{fig9}.
}
\end{figure}

\begin{figure}
\begin{center}
\includegraphics[width=0.85\columnwidth]{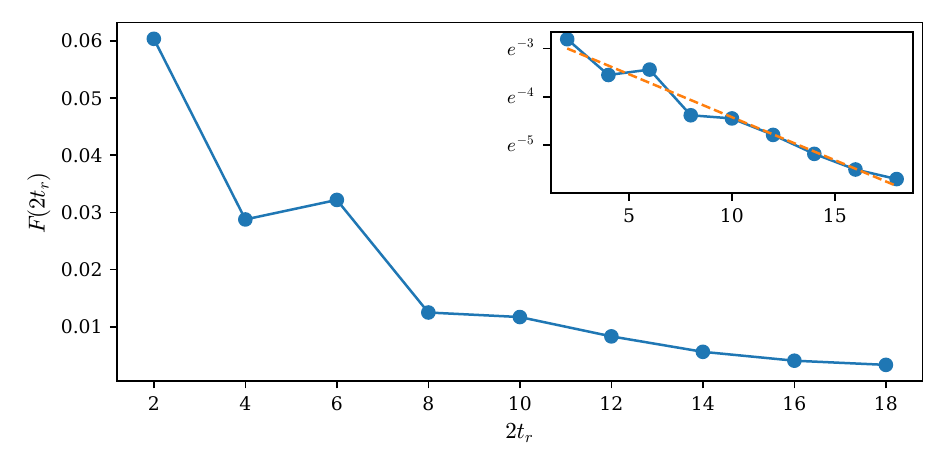}
\end{center}
\caption{\label{fig12}
  Fidelity $F(t=2t_r)$ vs time return moment $t=2t_r$
  for the butterfly effect of qubit when only one central qubit is inverted at time $t_r$,
  $n_q=5 \times 5 =25$, $\varepsilon = 2$, $h=1$,
  $|\psi_0\rangle=|\uparrow \uparrow \uparrow \dots \uparrow \rangle$ and
  there are no imperfections.
  Insert shows $\ln F(t=2t_r)$ vs $2t_r$ and a fitted dashed line gives
  $\ln F = -0.178 \times 2t_r - 2.64$ .
}
\end{figure}

The butterfly effect of qubit is shown in Fig.~\ref{fig9} and Fig.~\ref{fig10}
for fidelity $F(t)$ and entropy $S_q(t)$. Here the time reversal protocol
for the system with 25 qubits, without imperfections,
is applied at the time moment $t_r=30$ with inversion of central qubit
(located at the lattice position $x=y=3$). This one qubit inversion
breaks completely the time reversal: there are no signs of return
at $t=2t_r=60$ for $F$ and $S_q$. Also the time reversal is broken
for the system total spin squared $S^2(t)$ as it is shown in Fig.~\ref{fig12}
while in absence of qubit inversion $S^2(t)$ returned exactly to its initial value.

To illustrate the butterfly effect of qubit in a better way
we show in Fig.~\ref{fig12} the dependence of fidelity $F$ on the return time $2t_r$
when only one qubit is inverted at a time reversal moment $t_r$.
The results show that $F(2t_r)$ drops significantly already after one
evolution period with $t_r=1$. For $2t_r > 2$ the decay of
fidelity continues approximately exponentially
($F \sim \exp(-0.178 \times 2t_r)$ for parameters of Fig.~\ref{fig12} insert).
We note certain similarity of this decay with the results reported in \cite{arnold1}
(see Fig.3 there).

Thus we conclude that due to many-body quantum chaos
and related induced rapid entropy growth (see e.g. Fig.~\ref{fig5})
there is a strong butterfly effect of qubit
for the complex quantum evolution
of our system (\ref{hamil}) that completely breaks
time reversibility. We consider this butterfly effect of qubit
as an analog of the classical butterfly effect
discussed in \cite{lorenz1972,butterflywiki}.

\section{Discussion}
\label{sec7}

In this work we propose a simple quantum protocol
that allows one to perform time reversal of complex time evolution
on a quantum computer in a regime of quantum chaos.
In this protocol we consider a quantum computer qubits on a square lattice
with next-nearest neighbor static interactions of qubits.
The time evolution is considered for a short periodically pulsed global magnetic field
acting simultaneously on all qubits.
This corresponds to one-qubit phase rotation for all qubits.
The periodic time evolution is shown to be in a regime of quantum chaos
with the level spacing statistics of quasienergies
described by RMT. The time reversal is done by a change of period of pulses
and by inversion of magnetic field at a certain moment of time.
As a result the system exactly returns to its initial
quantum register state.
The time reversal is shown to be robust
against quantum imperfections.
The present abilities of quantum computers \cite{smela,smela2,91qu,144qu,74qu} with up to 144 qubits
or quantum annealers \cite{qa1,qa2,qa3} allow to perform this time reversal protocol
with about 100 or even larger number of qubits with
rather simple quantum operations. For a quantum simulator with 100 qubits a macroscopic
number of quantum trajectories in the Hilbert space of size $N_H \sim 10^{30}$
will be reversed in time returning back to their initial state.
This number is significantly larger than the Avogadro number $N_A \approx 6 \times 10^{23}$.

Even if the time reversal is robust with respect to gate imperfections
we show that the many-body quantum chaos leads to the butterfly effect of qubit
when inversion of only one qubit breaks time reversal of the whole system.

We argue that the experiments on quantum computers or quantum annealers will highlight
the Boltzmann-Loschmidt dispute within the quantum reality
150 years later.

\begin{acknowledgments}
The authors acknowledge support from the grant ANR France project NANOX $N^\circ$ ANR-17-EURE-0009 in the framework of the Programme Investissements d'Avenir (project MTDINA).
\end{acknowledgments}

\end{document}